\documentclass[floatfix,showkeys,aps]{revtex4}
\usepackage{graphicx}
\usepackage{color}
\usepackage{amsmath}
\usepackage{amssymb}
\usepackage{times}
\usepackage{units}
\usepackage{url}

\newcommand{\xit}{\xi_{\bot}}
\newcommand{\xia}{\xi_{\|}}

\begin{document}

\author{Tony S.\ Yu}
\author{Eric Lauga}
\author{A.\ E.\ Hosoi}
\affiliation{Hatsopoulos Microfluids Laboratory, Department of Mechanical Engineering, 
Massachusetts Institute of Technology, 
Cambridge, Massachusetts 02139}
\title{Experimental Investigations of Elastic Tail Propulsion at Low Reynolds Number}

\begin{abstract}
A simple way to generate propulsion at low Reynolds number is to periodically oscillate a passive flexible filament. 
Here we present a macroscopic experimental investigation of such a propulsive mechanism.
A  robotic swimmer is constructed and both tail shape and propulsive force are measured.
Filament characteristics and the actuation are varied and resulting data are quantitatively compared with existing linear and nonlinear theories.  
\end{abstract}

\maketitle

At small scales, the physics of swimming is fundamentally different than at mesoscopic scales as the dominance of viscous forces over inertial forces leads to equations of motion that are time-reversible. In his famous lecture, \textit{Life at Low Reynolds Numbers}, Purcell \cite{Purcell1977} described three simple swimming mechanisms that are not time-reversible and hence lead to a net translation in the absence of inertial effects: (1) the ``corkscrew'' \cite{Purcell1997}, in which a rigid helical filament is rotated in a viscous  liquid, analogous to the swimming mechanism of many bacteria  \cite{bergbook,braybook}; (2) the ``three-link swimmer,'' the simplest rigid-linked mechanism that swims without inertia \cite{Becker2003}; and (3) the ``flexible oar'' \cite{Wiggins1998,Wiggins1998a,Lowe2003}, in which a flexible tail is oscillated in a viscous fluid, generating traveling waves along the filament that produce a propulsive force (see also \cite{machin58,machin63,kim06}).  The purpose of this letter is to experimentally investigate  the flexible oar design and to compare the resulting force data with existing theories. 

Swimming at micro-scales has long been the realm of bacteria and other microorganisms \cite{braybook,lighthill76} but contemporary advances have allowed engineers to catch up with nature. Dreyfus {\sl et al.} have recently created the first manmade micro-swimmer  \cite{Dreyfus2005}, in which a chain of paramagnetic beads propagates a bending wave along the chain driven by an external magnetic field. Although construction of this remarkable swimmer was at least partially motivated by existing flexible tail theories  \cite{Wiggins1998,Wiggins1998a,Lowe2003,machin58,machin63,kim06}, the mechanism is not a truly passive flexible tail as internal torques are applied along the length of the filament.  A second experiment performed by Wiggins {\sl et al.} measured the shape changes of a passive actin filament, oscillated at one end via optical tweezers \cite{Wiggins1998a}.  The shapes recorded in these trials match elastohydrodynamic theory well however the resulting  propulsive force -- a key parameter in designing microscopic swimmers -- was not measured.
Here we propose the first experimental determination of this force and show that the linear theory due to Wiggins and Goldstein  \cite{Wiggins1998} quantitatively predicts both the shape of the elastic filament and the resulting propulsive, viscous forces.

\begin{figure}[t]
	\centering
	\includegraphics{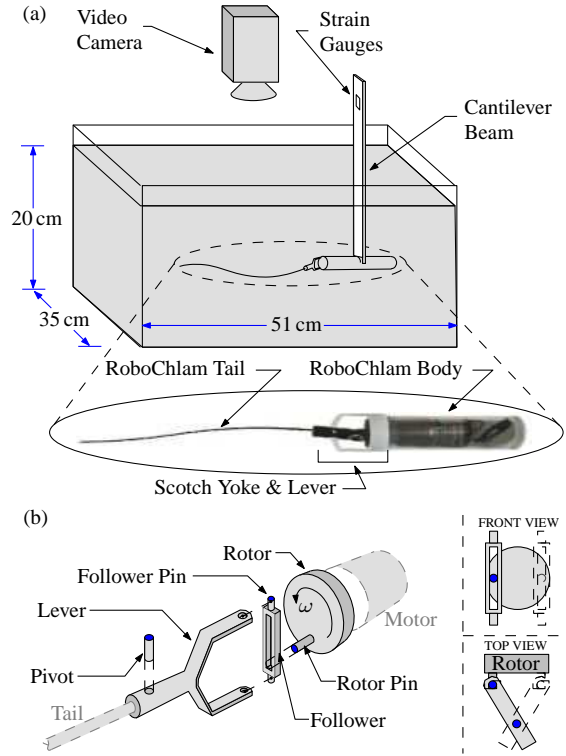}
	\caption{(a) Experimental setup to measure tail shapes and propulsive forces. (b) Scotch yoke and lever mechanism. The rotor and follower form the Scotch yoke, which converts the motor's rotation into a translational oscillation. This oscillation is then converted to an angular oscillation by a lever. The angular oscillation is approximately sinusoidal for a constant motor rotation.}
	\label{fig:ExperimentalSetup}
\end{figure}

In order to experimentally quantify the propulsive characteristics of the flexible oar design, we built a robotic swimmer dubbed ``RoboChlam'' (after the algae \textit{Chlamydomonas}), as is displayed in Fig.~\ref{fig:ExperimentalSetup}(a). The RoboChlam body was approximately \unit[8]{cm} in length and housed a geared DC motor. The motor's rotation was converted into an angular oscillation using a Scotch yoke and a lever (see Fig.~\ref{fig:ExperimentalSetup}b). Consequently, the tail was angularly-actuated: the base of the filament was fixed at the origin and the base-angle was varied sinusoidally with an amplitude $a_0$ and a frequency $\omega$. The voltage across the motor governed the oscillation frequency (between $5$ and $\unit[0.4]{rad/s}$), and the length of the lever controlled the amplitude of oscillation ($\unit[0.814]{rad}$ and $\unit[0.435]{rad}$). At the end of the lever, stainless steel wires of length $\unit[18]{cm}$ to $\unit[30]{cm}$ acted as elastic tails. Two different tail diameters were used in these experiments: $D = \unit[0.5]{mm}\;\mathrm{and}\;\unit[0.61]{mm}$ resulting in bending stiffnesses of $6.1\times 10^{-4}\;\mathrm{and}\;\unit[1.3\times 10^{-3}]{N\cdot m^2}$, respectively.

\begin{figure}[t]
	\centering
	\includegraphics{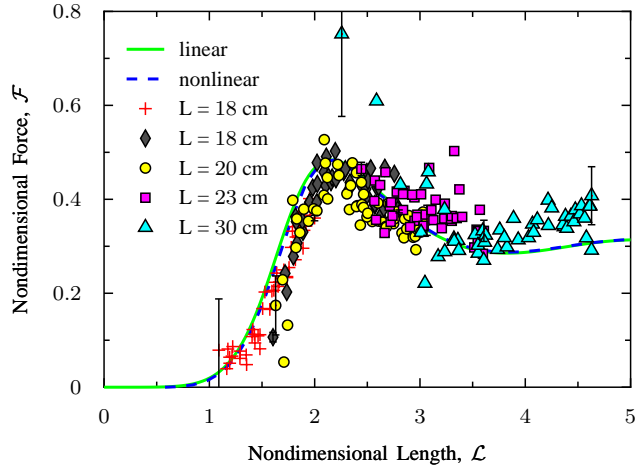}
	\caption{Force measurements for various tail lengths, $L$. Oscillation frequency was varied to span a range of dimensionless lengths, $\mathcal{L}$ where the dimensionless length and force are defined in equations (\ref{DimensionlessLength}) and (\ref{eq:NondimForce}) respectively. The \textcolor{red}{+} symbols correspond to $D = \unit[0.61]{mm}$ and $a_0 = \unit[0.814]{rad}$. All other data correspond to  $D = \unit[0.5]{mm}$ and $a_0 = \unit[0.435]{rad}$.  There are no free parameters in the comparison between experiment and theory.}
	\label{fig:ForcePlot}
\end{figure}

\begin{figure}[t]
	\centering
 	\includegraphics{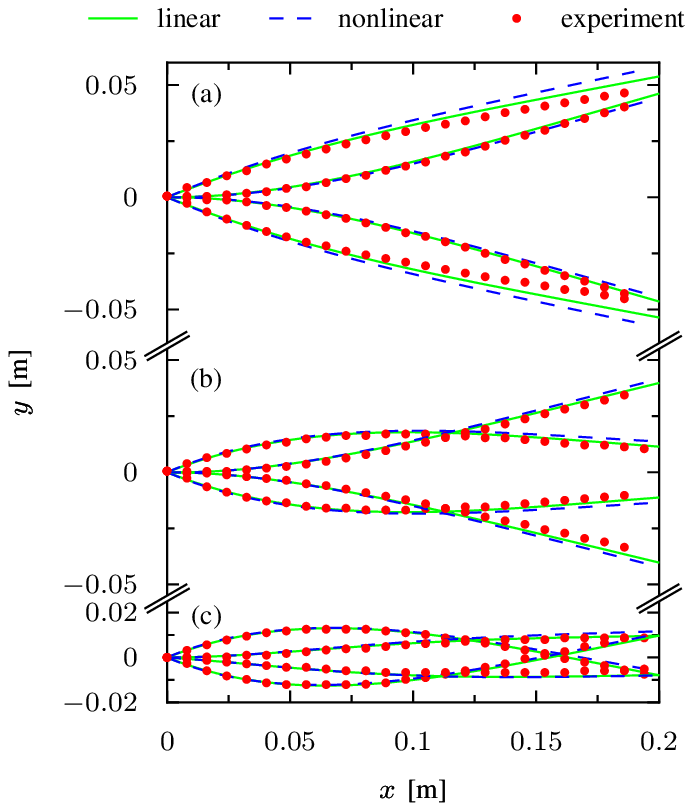}
	\caption{Comparison between experiment, linear and nonlinear theories of tail shapes.  Snapshots are shown at four points in the cycle for one tail with $L = \unit[20]{cm}$, $D = \unit[0.5]{mm}$, $a_0 = \unit[0.435]{rad}$, at three different oscillation frequencies: (a) $\omega = \unit[0.50]{rad/s}$ ($\mathcal{L} = 1.73$), (b) $\omega = \unit[1.31]{rad/s}$ ($\mathcal{L} = 2.20$), (c) $\omega = \unit[5.24]{rad/s}$ ($\mathcal{L} = 3.11$).}
	\label{fig:TailShapes}
\end{figure}

RoboChlam was immersed in high viscosity ($\unit[3.18]{Pa\cdot s}$) silicone oil to approach the low Reynolds numbers ($10^{-2}$ to $10^{-3}$) achieved by microorganisms. 
Tail shapes generated by RoboChlam were imaged with a video camera at 30 frames per second and 720 $\times$ 480 pixels per frame. A cantilever beam anchored the device, and a pair of strain gauges on opposite sides of the beam measured beam deflection. Strain gauge readings were converted into force measurements; a no-load voltage reading was taken at the beginning and end of each trial to measure the thermal drift in the strain gauges and the accompanying circuitry. Although force data were obtained through measured deflections of the cantilever beam, this deflection was small -- less than half a centimeter at the beam's tip -- thus, RoboChlam's position was approximately fixed. Experiments showed that an angular oscillation starting with the tail at rest reached steady-state motion after approximately two periods of oscillation; the time scale associated with this decay of transients corresponds well with the transient time scales observed in our nonlinear simulations. Finally, videos of the tail shapes were digitized for comparison to simulations and theoretical predictions.   Experimental data are summarized in Figs.~\ref{fig:ForcePlot} and \ref{fig:TailShapes}.

\begin{figure}[t]
	\centering
	\includegraphics{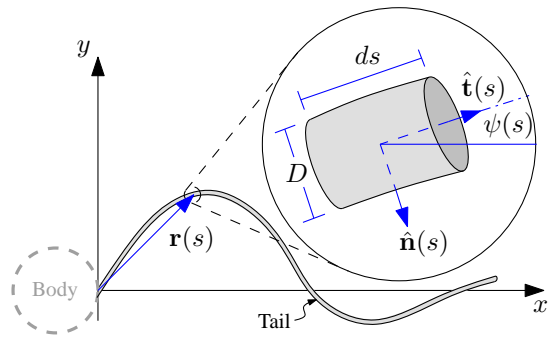}
	\caption{Schematic of the elastic tail with the origin defined at the base of the tail.}
	\label{fig:TailDiagram}
\end{figure}

To perform a quantitative comparison of experimental force and shape data with theoretical predictions, we first briefly review a few key results of the theory of actuated elastic filaments in Stokes flow \cite{Wiggins1998a}. Consider an elastic, cylindrical rod whose base is attached to a fixed body (see Fig.~\ref{fig:TailDiagram}). In a low Reynolds number regime, the inertia of the fluid can be neglected and the fluid dynamics is well-described by Stokes equations. If the length of the tail, $L$, is much greater than its diameter, $D$, the hydrodynamics can be further simplified by using slender body theory, the approximation of which is resistive force theory \cite{gray55,lighthill76,brennen77}. Thus, the drag forces on the tail are linearly related to the velocity through the transverse and axial drag coefficients, $\xit$ and $\xia$, respectively, and the drag force per unit length of the rod can be expressed as
\begin{equation}\label{eq:DragForce}
    \mathbf{f}_d = - [\xit\hat{\mathbf{n}}\hat{\mathbf{n}} + \xia\hat{\mathbf{t}}\hat{\mathbf{t}}] \cdot \mathbf{r}_t
\end{equation}
where the subscript $t$ denotes a derivative in time, $\mathbf{r}$ is the position vector of a point along the tail, and $\hat{\mathbf{n}}$ and $\hat{\mathbf{t}}$ are the unit normal and tangent to the filament, respectively. We consider in this letter a planar actuation of the rod, so that $\hat{\mathbf{n}}$ is defined without ambiguities to remain in this plane.

The elastic forces on the rod are derived from an energy functional which includes bending energy and an inextensibility constraint
\begin{equation}
\mathcal{E} = \int_0^L \left[\frac{A}{2}\kappa^2 + \frac{\Lambda}{2} {\mathbf{r}_s}^{2} \right]\,{\rm d }s
\end{equation}
where $A$ is the bending stiffness, $\kappa$ is the local curvature of the tail and $\Lambda$ is the Lagrange multiplier enforcing inextensibility. Using calculus of variation  we obtain the elastic force per unit length, $ \mathbf{f_{\epsilon}} = -\delta  {\cal E} /\delta \mathbf{r} $ 
as given by  \cite{Camalet2000,wolgemuth00}
\begin{equation}\label{eq:ElasticForce}
	\mathbf{f_{\epsilon}}= -(A\psi_{sss} - \psi_s\tau)\hat{\mathbf{n}} + (A\psi_{ss}\psi_s + \tau_s)\hat{\mathbf{t}}
\end{equation}
where the subscript $s$ denotes a derivative in the coordinate along the tail axis, $\psi$ is the local angle (see Fig.~\ref{fig:TailDiagram}), and $\tau$ can also be interpreted as the local tension in the tail.

Local mechanical equilibrium along the rod, leads to  a pair of coupled, nonlinear partial differential equations of motion, as shown in \cite{Camalet2000}:
\begin{eqnarray}
        \psi_t &= &- \frac{1}{\xit} \left(A\psi_{ssss} - \tau\psi_{ss} - \tau_s\psi_s\right)  \label{eq:PsiEoM}\\
        	&&+ \frac{1}{\xia} \left(A{\psi_s}^2\psi_{ss} + \tau_s\psi_{s}\right), \nonumber \\
	\tau_{ss} - \frac{\xia}{\xit}\tau{\psi_s}^2 &= &
				- A\frac{\xit-\xia}{\xit}\psi_s\psi_{sss} - A{\psi_{ss}}^2. \label{eq:TauEoM}
\end{eqnarray} 
Numerical solutions to these equations were found using a Newton-Raphson iteration and are plotted along with experimental data in Figs.~\ref{fig:ForcePlot} and \ref{fig:TailShapes}.

For small deflections (\textit{i.e.} assuming $\psi \ll 1$ such that $\psi \approx y_x$),
Wiggins and Goldstein \cite{Wiggins1998} have shown that the motion of the tail can be further simplified and is described by a linear, ``hyperdiffusion'' equation:
\begin{equation}\label{eq:LinearEoM}
    y_t \approx -\frac{A}{\xit}y_{xxxx}
\end{equation}
where subscripts $x$ and $t$ denote derivatives in position and time, respectively. 
For the case of harmonic angular-actuation, we apply the boundary condition  $\psi = a_0 \sin(\omega t)$ at the base. 
The nondimensionalization of Eq.~\eqref{eq:LinearEoM} is obtained by substituting $x = L\tilde{x}$, $y = a_0 L\tilde{y}$, and $t = \tilde{t}/\omega$ into Eq.~(\ref{eq:LinearEoM}), leading to $\tilde{y}_{\tilde{t}} \approx -(\ell_{\omega}/ L)^4 \tilde{y}_{\tilde{x}\tilde{x}\tilde{x}\tilde{x}}$ where, $\ell_{\omega} = (A/\omega\xit)^{1/4}$, is the characteristic penetration length of the elastohydrodynamic problem; solutions to Eq.~\eqref{eq:LinearEoM} decay exponentially in space over this typical length scale. The time-evolution of the tail shapes is then only a function of the angular amplitude, $a_0$, and the dimensionless length, 
\begin{equation}
\mathcal{L} = L/\ell_{\omega} = L\left( \frac{\omega \xit}{A}\right)^{1/4}.
\label{DimensionlessLength}
\end{equation}
This dimensionless length is the key parameter in the problem and represents the ``floppiness" of the tail and hence the overall effectiveness of the swimmer.  In particular, theory predicts an optimal dimensionless tail length as both short, stiff tails and long, flexible tails produce negligible net translation -- the first is ineffective owing to the scallop theorem and the second owing to the excessive drag on the long passive filament. 

For a tail that is periodically oscillated, Eq.~(\ref{eq:LinearEoM}) can be solved analytically \cite{Wiggins1998, Wiggins1998a}.
At the base of the filament, the reaction forces and torque must balance the drag forces along the tail. The opposite end of the tail is force and torque-free such that $\psi_s = 0$, $\psi_{ss} = 0$, and $\tau = 0$ at $s = L$. For small deflections, the $x$-component of local drag force, Eq.~(\ref{eq:DragForce}), can be integrated along the length of the tail to yield the propulsive force
\begin{equation}\label{eq:LinearForce}
	\langle F \rangle \approx -A \frac{\xit-\xia}{\xit} \langle y_x y_{xxx}- \frac{1}{2} {y_{xx}}^2 \rangle_{x = 0}
\end{equation}
where $\langle\ldots\rangle$ denotes averaging over one period of oscillation. Note that Eq.~(\ref{eq:LinearForce}) differs from the one presented in \cite{Wiggins1998,Wiggins1998a} by a factor $(\xit-\xia)/\xit$; this disparity arises from a proper integration of the drag force on the filament \cite{Lowe2003}.

In order to obtain results valid beyond the small-slope approximation, our numerical solutions to the full nonlinear system (Eqs.~\ref{eq:PsiEoM} and \ref{eq:TauEoM}) were employed and the propulsive force was found by numerical integration of the local drag force along the length of the tail.  A dimensionless force $\mathcal{F}$ was defined by substituting $x = \ell_{\omega} \tilde{x}$, $y = a_0\ell_{\omega}\tilde{y}$, and $t = 2\pi\tilde{t}/\omega$ into Eq.~(\ref{eq:LinearForce}), such that
\begin{equation}\label{eq:NondimForce}
    \langle F \rangle = {a_0}^2{\ell_{\omega}}^2 (\xit-\xia)|\omega| \langle\mathcal{F}\rangle.
\end{equation}
Since the distance from the tail to the nearest wall was on the order of the tail's length, drag coefficients corrected for wall effects as in \cite{Demestre1975} were used in simulations and for nondimensionalizing the force data. These wall corrections have been shown to match well with experimental results \cite{Stalnaker1979}; for simplicity, the effect of only a single side-wall was considered. These equations produced a drag difference of approximately $\xit - \xia = \unit[3.35]{Pa\cdot s}$ -- about $40\%$ greater than the drag difference without wall effects.

\begin{figure}[t]
	\centering
	\includegraphics{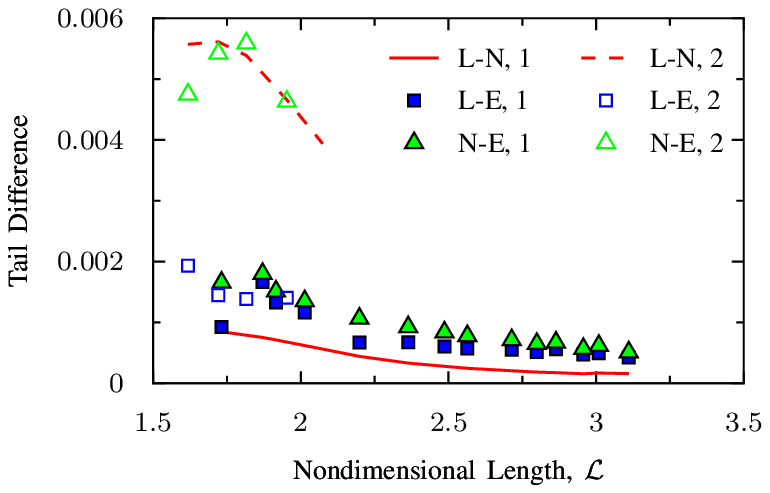}
	\caption{Normalized, time-averaged differences between linear (L), nonlinear (N), and experimental (E) tail shapes. The difference is calculated as the $\ell^2$-norm of the vertical-distance vector between two tails divided by the tail length and the number of points along the tail. Two data sets are shown: (1) $L = \unit[20]{cm}$, $D = \unit[0.5]{mm}$, and $a_0 = \unit[0.435]{rad}$; (2) $L = \unit[18]{cm}$, $D = \unit[0.63]{mm}$, and $a_0 = \unit[0.814]{rad}$.}
	\label{fig:TailDifferences}
\end{figure}

The results of our investigations are summarized in Figs.~\ref{fig:ForcePlot}, \ref{fig:TailShapes} and \ref{fig:TailDifferences}.
We first display in Fig.~\ref{fig:ForcePlot} the propulsive force generated for a range of dimensionless tail lengths, $\mathcal{L}$. All parameters of the experiment were known or measured, and no fitting of data was necessary. We obtain excellent agreement of the propulsive force with the theoretical (linear model, Eq.~\ref{eq:LinearEoM}) and numerical values  (nonlinear model, Eqs.~\ref{eq:PsiEoM} and \ref{eq:TauEoM}). The force data from the RoboChlam experiments show a maximum dimensionless force at $\mathcal{L} \approx 2.1$, in agreement with prediction from the theory. Note that our data was nondimensionalized with the drag difference, $\xit-\xia$ (see Eq.~\ref{eq:NondimForce}), instead of the transverse drag $\xit$, which was used in \cite{Wiggins1998, Wiggins1998a}. The drag difference orginated in Eq.~(\ref{eq:LinearForce}), and it represents the correct scaling as a tail with isotropic drag ($\xit = \xia$) should produce zero propulsive force \cite{Becker2003,Lowe2003}. We note also that the maximum value of $\mathcal{L}$ that could be tested was limited by the motor's rotation rate and the length of tail that would fit in the experimental apparatus.

In comparing the data to linear elastohydrodynamic theories, there are three primary sources of error: wall effects, thermal drift in the experiment, and the neglected nonlinearities in the theory.
The error bars in Fig.~\ref{fig:ForcePlot} arise from uncertainty in the no-load voltage of the strain gauge measurements. At lower oscillation frequencies, the sample time of the experiment increased, leading to noticeable thermal drift in strain gauge (force) measurements and thus, larger drift error for the left-most points of a given data-set. In addition, the tip of the longest tail ($\unit[30]{cm}$, $\color{cyan}{\blacktriangle}$) was only a few centimeters from the back wall and thus, this wall had a non-negligible effect on the drag of the longest tail resulting in an increased thrust as expected.  Recall that our wall-corrected drag coefficients only account for a single wall -- the side wall rather than the back wall -- of the tank, as is appropriate for all but the longest tails in our experiments.  It is interesting to note that, in these experiments,  nonlinear effects are completely negligible relative to the other two sources of error even for long tails and large actuation angles. 
 
In Fig.~\ref{fig:TailShapes} we plot the tail shapes from experiments along with simulationed shapes from both the linear and nonlinear theories. The plot shows three tails from a single data set (constant $L$, $D$, and $a_0$, but varying $\omega$) with dimensionless lengths (a) $\mathcal{L} = 1.73$, (b) $\mathcal{L} = 2.20$, and (c) $\mathcal{L} = 3.11$. These dimensionless lengths span the region near the maximum dimensionless force. The tail shapes from experiment matched well with those from the linear and nonlinear simulations, and only slight differences between the three tails were observed. Tails whose dimensionless length was small (Fig.~\ref{fig:TailShapes}a) moved stiffly, while those with large dimensionless lengths (Fig.~\ref{fig:TailShapes}c) were flexible, as predicted by theory. The difference between the different tail shapes (theory, experiments, simulations) is quantified in Fig.~\ref{fig:TailDifferences}. The measured errors are observed to be small. The fact that the data match the linear simulation better than the nonlinear solution is fortuitous and merely reflects the fact that resistive force theory is only an approximation of the equation of hydrodynamics \cite{lighthill76}. 

In summary, we have presented an experimental investigation of Purcell's flexible oar swimmer. Measurements of propulsive forces and time-varying shapes are in agreement  with the results of resistive-force theory. Remarkably, the small-slope model of Wiggins and Goldstein \cite{Wiggins1998} appears to remain quantitatively correct well beyond its regime of strict validity.


Our future work will investigate the efficiency of this propulsive mechanism when embedded in a synthetic free-swimmmer -- that is, an elastic filament  attached to a body which translates and rotates with the forces and torque generated by the propulsive tail. Preliminary free swimming experiments show that rotation of the swimmer body significantly changes the shapes of the tail, modifying the force curve shown in Fig.\ \ref{fig:ForcePlot}, and appreciably impacting the dynamics of the swimmer.

\bibliography{ElasticTail}

\begin{thebibliography}{10}

\bibitem{Purcell1977}
E.~M. Purcell.
\newblock Life at low reynolds number.
\newblock {\em Am. J. Phys.}, 45:3--11, 1977.

\bibitem{Purcell1997}
E.~M. Purcell.
\newblock The efficiency of propulsion by a rotating flagellum.
\newblock {\em P. Natl. Acad. Sci. USA}, 94:11307--11311, 1997.

\bibitem{bergbook}
H.~C. Berg.
\newblock {\em {\it E. coli} in {M}otion}.
\newblock Springer-Verlag, NY, 2004.

\bibitem{braybook}
D.~Bray.
\newblock {\em Cell Movements}.
\newblock Garland Publishing, New York, NY, 2000.

\bibitem{Becker2003}
L.~E. Becker, S.~A. Koehler, and H.~A. Stone.
\newblock On self-propulsion of micro-machines at low reynolds number:
  Purcell's three-link swimmer.
\newblock {\em J. Fluid Mech.}, 490:15 -- 35, 2003.

\bibitem{Wiggins1998}
C.~H. Wiggins and R.~E. Goldstein.
\newblock Flexive and propulsive dynamics of elastica at low reynolds number.
\newblock {\em Phys. Rev. Lett.}, 80:3879 -- 3882, 1998.

\bibitem{Wiggins1998a}
C.~H. Wiggins, D.~Riveline, A.~Ott, and R.~E. Goldstein.
\newblock Trapping and wiggling: Elastohydrodynamics of driven microfilaments.
\newblock {\em Biophys. J.}, 74:1043--1060, 1998.

\bibitem{Lowe2003}
C.~P. Lowe.
\newblock Dynamics of filaments: modelling the dynamics of driven
  microfilaments.
\newblock {\em Philos. Trans. R. Soc. London, Ser. B}, 358:1543--1550, 2003.

\bibitem{machin58}
K.~E. Machin.
\newblock Wave propagation along flagella.
\newblock {\em J. Exp. Biol}, 35:796--806, 1958.

\bibitem{machin63}
K.~E. Machin.
\newblock The control and synchronization of flagellar movement.
\newblock {\em Proc. Roy. Soc. B}, 158:88--104, 1963.

\bibitem{kim06}
Y.~W. Kim and R.~R. Netz.
\newblock Pumping fluids with periodically beating grafted elastic filaments.
\newblock {\em Phys. Rev. Lett.}, 96, 2006.

\bibitem{lighthill76}
J.~Lighthill.
\newblock Flagellar hydrodynamics - {The John von Neumann} lecture, 1975.
\newblock {\em SIAM Rev.}, 18:161--230, 1976.

\bibitem{Dreyfus2005}
R.~Dreyfus, J.~Baudry, M.~L. Roper, M.~Fermigier, H.~A. Stone, and J.~Bibette.
\newblock Microscopic artificial swimmers.
\newblock {\em Nature}, 437:862--865, 2005.

\bibitem{gray55}
J.~Gray and G.~J. Hancock.
\newblock The propulsion of sea-urchin spermatozoa.
\newblock {\em J. Exp. Biol.}, 32:802--814, 1955.

\bibitem{brennen77}
C.~Brennen and H.~Winet.
\newblock Fluid mechanics of propulsion by cilia and flagella.
\newblock {\em Ann. Rev. Fluid Mech.}, 9:339--398, 1977.

\bibitem{Camalet2000}
S.~Camalet and F.~Julicher.
\newblock Generic aspects of axonemal beating.
\newblock {\em New J. Phys.}, 2:1--23, 2000.

\bibitem{wolgemuth00}
C.~W. Wolgemuth, T.~R. Powers, and R.~E. Goldstein.
\newblock Twirling and whirling: {Viscous} dynamics of rotating elastic
  filaments.
\newblock {\em Phys. Rev. Lett.}, 84:1623--1626, 2000.

\bibitem{Demestre1975}
N.~J. De~Mestre and W.~B. Russel.
\newblock Low-reynolds-number translation of a slender cylinder near a plane
  wall.
\newblock {\em J. Eng. Math.}, 9:81--91, 1975.

\bibitem{Stalnaker1979}
J.~F. Stalnaker and R.~G. Hussey.
\newblock Wall effects on cylinder drag at low reynolds number.
\newblock {\em Phys. Fluids}, 22:603--613, 1979.

\end{thebibliography}
\bibliographystyle{unsrt}

\end{document}